\documentclass{webofc}

\usepackage[varg]{txfonts}   
\usepackage{hyperref}
\usepackage{url}
\hypersetup{colorlinks=true,citecolor=blue,urlcolor=blue,linkcolor=blue}
\def\kompost{K\o{}MP\o{}ST }

\begin{document}
\title{Dilepton emission in heavy ion collisions and chemical equilibrium of QCD matter}
%
%

\author{\firstname{Xiang-Yu} \lastname{Wu}\inst{1}\fnsep\thanks{\email{xiangyu.wu2@mail.mcgill.ca}} \and
        \firstname{Lipei} \lastname{Du}\inst{1,2,3}\fnsep\thanks{\email{lipei.du@mail.mcgill.ca}} \and
        \firstname{Charles} \lastname{Gale}\inst{1}\fnsep\thanks{\email{charles.gale@mcgill.ca }} \and
        \firstname{Sangyong} \lastname{Jeon}\inst{1}\fnsep\thanks{\email{sangyong.jeon@mcgill.ca}}
        }
        
\institute{Department of Physics, McGill University, Montreal, Quebec, Canada H3A 2T8 \and
Department of Physics, University of California, Berkeley, California 94270, USA \and 
Nuclear Science Division, Lawrence Berkeley National Laboratory, Berkeley, California 94270, USA
}

\abstract{
We study thermal dilepton production and anisotropic flow in Pb+Pb collisions at $\sqrt{s_{NN}} = 5.02 \, \mathrm{TeV}$ using next-to-leading-order (NLO) thermal QCD dilepton emission rates. A hybrid model (IP-Glasma+\kompost+MUSIC+UrQMD) simulates the collision evolution. The role of the pre-equilibrium stage in dilepton observables is examined. We also explore how chemical equilibrium in QCD matter affect dilepton observables.

}
\maketitle
\section{Introduction}
\label{intro}
Dilepton production is one of the clean probes
for studying
the properties of quark-gluon plasma (QGP) formed in relativistic heavy-ion collisions at the Relativistic Heavy Ion Collider (RHIC) and at the Large Hadron Collider (LHC).
Since dileptons are emitted 
from
all stages of the heavy ion collision,
and can escape the QGP medium without further 
interactions, 
they complement final-state soft hadrons. While hadrons  generally  carry information  about the QGP matter near the freeze-out temperature,  dileptons can directly probe the early stage of the QGP fireball, where the typical temperature is  higher than the freeze-out temperature.
In addition, because dileptons possess an additional degree of freedom, the invariant mass, which is not influenced by the radial flow of the QGP medium, they are often considered
a good thermometer for
QGP\cite{Churchill:2023zkk}.

As a common strategy, heavy-ion collision modeling begins with an initial state, 
followed by hydrodynamic evolution, and a hadronic scattering stage. This multistage modeling approach has achieved significant success in describing collective flow phenomena and in extracting key transport properties of the QGP medium, such as shear and bulk viscosities\cite{JETSCAPE:2020shq}. 
Recently, in addition to this widely adopted standard multi-stage model,  attention has been paid to the dynamical evolution prior to the hydrodynamic stage, namely the pre-equilibrium stage. It has been found that the strongly anisotropic momentum distribution of quarks \cite{Garcia-Montero:2024lbl,Garcia-Montero:2023lrd,Kasmaei:2018oag} and the initially gluon-dominated medium in the pre-equilibrium stage can affect the results in the soft sector and for electromagnetic (EM) probes (dilepton and photon).
In addition, EM probes provide an excellent opportunity to investigate chemical equilibrium process
in the early stages of heavy-ion collisions\cite{Gale:2021emg,Wu:2024pba}. This is because the production of dileptons and photons is highly sensitive to the net quark content within a given space-time cell, which gradually evolves during the pre-equilibrium stage. 

In this work, we investigate the impact of the pre-equilibrium stage dynamics and chemical equilibration on thermal dilepton production and anisotropic flow in Pb+Pb collisions at LHC an energy ($\sqrt{s_{NN}}$ = 5.02 TeV) by combining the multi-stage hydrodynamic framework \cite{Schenke:2010nt,Schenke:2010rr,Paquet:2015lta,Kurkela:2018wud,Kurkela:2018vqr} with the NLO thermal QCD  dilepton emission rates\cite{Churchill:2023vpt}.

\section{Multi-stage hydrodynamic framework}
\label{framework}
We simulate the dynamical evolution of the QGP medium using the state-of-the-art iEBE-MUSIC hybrid framework \cite{Schenke:2010nt,Schenke:2010rr,Paquet:2015lta}. The initial conditions are generated with (2+1)-D IP-Glasma \cite{Schenke:2012wb}, which describes the collision of two color glass condensates.  The classical Yang-Mills equations are solved to obtain the non-equilibrium gluon-dominated initial state. Then, during the pre-equilibrium stage, the system’s evolution is simulated with the \kompost framework\cite{Kurkela:2018wud,Kurkela:2018vqr}. Here, the energy-momentum tensor $T^{\mu\nu}$ is decomposed into two components: (1) a locally homogeneous background obeying universal scaling laws from the kinetic theory and (2) perturbations calculated via non-equilibrium linear response theory. This approach bridges the gap between the initial IP-Glasma states and the onset of hydrodynamic behavior. The subsequent hydrodynamic evolution solves energy-momentum conservation equations with second-order Israel-Stewart diffusion equations, including shear viscosity ($\eta/s$ = 0.12) and temperature-dependent bulk viscosity ($\zeta(T)$). The equation of state (EoS) is the lattice QCD calculation of the HotQCD collaboration\cite{HotQCD:2014kol}. When the local energy density drops below $\epsilon_{\rm frz}$ = 0.18 GeV/fm$^3$, we apply the Cooper-Frye prescription to sample hadrons, followed by hadronic rescattering and decay processes simulated using the UrQMD transport model. 
For model calibration, we used experimental data from the ALICE collaboration\cite{ALICE:2019hno,ALICE:2016ccg}, including multiplicity $ \frac{dN}{dy} $, mean transverse momentum $\langle p_T \rangle $ of identified hadrons and anisotropic flow coefficients $ v_n\{2\}$. The comparison of model with data is good in central and semi-central collisions, but discrepancies emerge in peripheral collisions\cite{Wu:2024pba,Gale:2021emg,Schenke:2020mbo}. These deviations stem from the vanishing bulk viscosity in the IP-Glasma and \kompost models, due to the assumed conformal symmetry. For further details on the model and hadron results, see Ref.\cite{Wu:2024pba}

\section{Dilepton emission rate and chemical equilibrium}\label{dilepton_rate}
The fully differential thermal dilepton emission rate $\Gamma_{ee}$ in the local fluid rest frame is expressed as
\begin{equation}
\frac{d \Gamma_{ee}}{d^4 P}=\frac{2 \alpha_{\rm em}^2 f_B(E)}{9 \pi^3 M^2} B\left(\frac{m_e^2}{M^2}\right) \rho_V(E,\boldsymbol{P})\,,
\end{equation}
where $P^{\mu}=(E,\boldsymbol{P})$ denotes the four-momentum of lepton pairs, $f_B(E) $  is the Bose-Einstein distribution function and $M=\sqrt{E^2-\boldsymbol{P}^2}$ represents the invariant mass. The number of quark flavors is taken here as 3. The function $B\left(\frac{m_e^2}{M^2}\right)$ represents a kinematic factor, while $\rho_V(E,\boldsymbol{P})$ denotes the vector spectral function. The NLO corrections in $\rho_V(E,\boldsymbol{P})$ include contributions from two-loop diagrams and the Landau-Pomeranchuk-Migdal (LPM) effect. This dilepton rate can also be extended to finite chemical potential regions\cite{Churchill:2023vpt}. Once the differential thermal dilepton rate is determined, the total thermal dilepton spectra are obtained by integrating over all space-time cells with local temperature $T$  and flow velocity $u^{\mu}$. The anisotropic flow coefficients of thermal dileptons are estimated using the scalar product method via a correlation with the hadronic reference flow\cite{Vujanovic:2019yih,Vujanovic:2017psb,Gale:2021emg}.

In this work, thermal dilepton production is attributed to two distinct sources: one from the hydrodynamic phase and the other from the pre-equilibrium phase. Since the current \kompost model describes a gluon-dominated system, we introduce an effective suppression factor ${\rm SF}(T,\tau)$ to dynamically model fermion production during the pre-equilibrium stage, corresponding to the gradual establishment of chemical equilibrium\cite{Gale:2021emg,Kurkela:2018xxd}.  Finally, we also consider Drell-Yan (pQCD) dileptons calculated using the DYturbo package \cite{Camarda:2019zyx}, which provides NLO pQCD predictions, given nuclear parton distribution functions.

\section{Results}
\begin{figure}[h]
\centering
\includegraphics[width=0.515\textwidth]{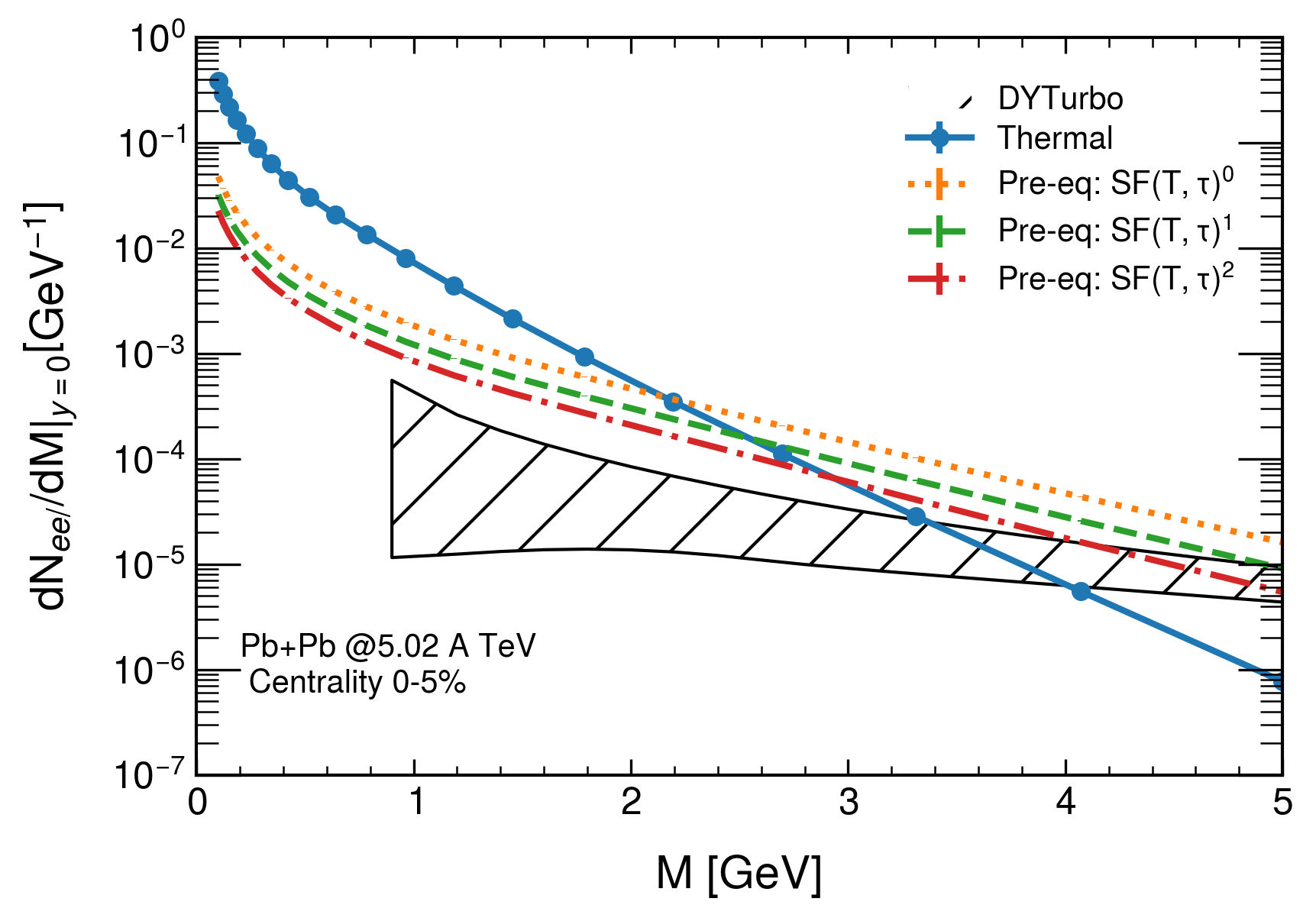}
\includegraphics[width=0.45\textwidth]{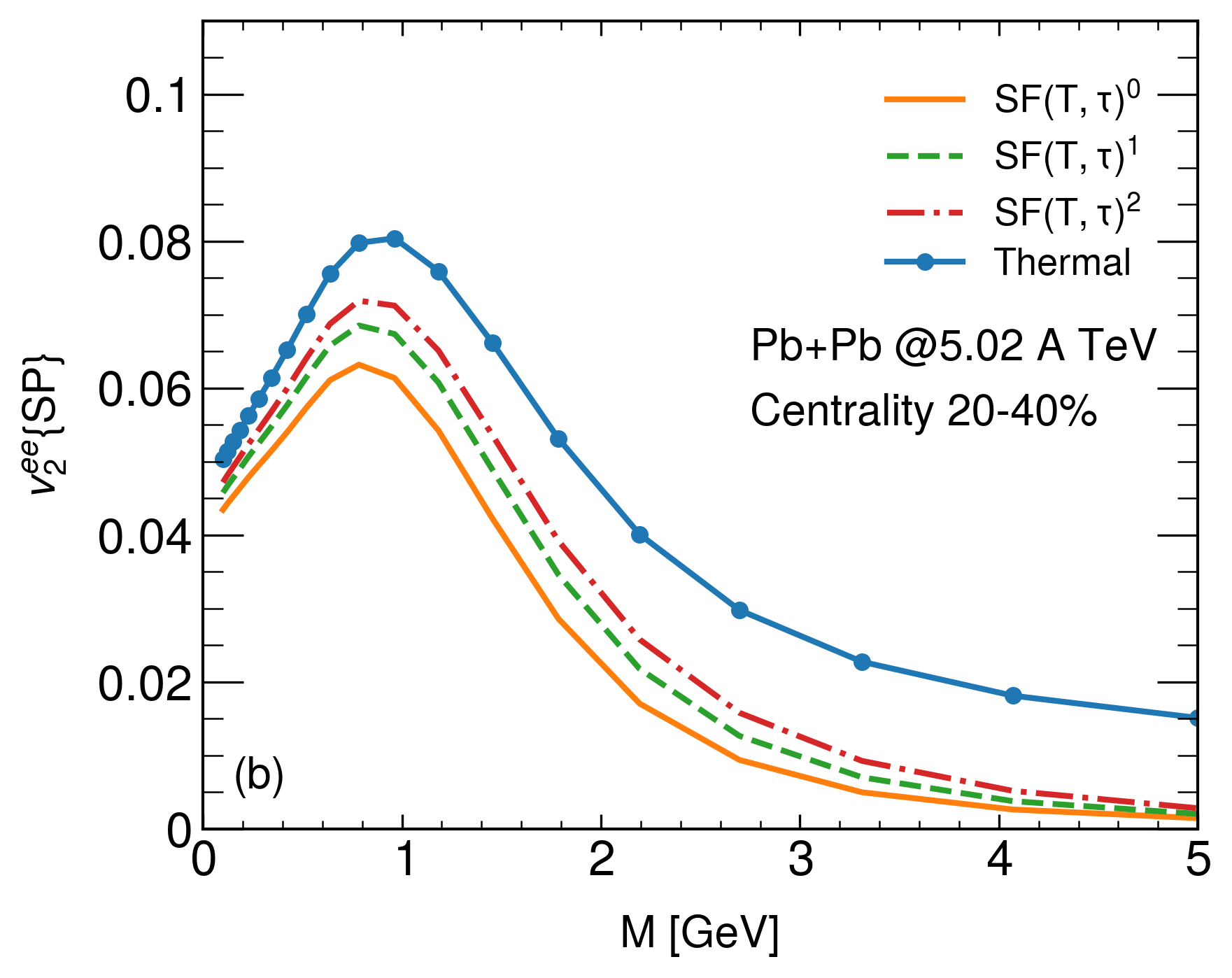}
\caption{Dilepton production yield (left panel) in the 0--5\% centrality bin and dilepton elliptic flow $v_2^{ee}\{\rm SP\}$ (right panel) in the 20--40\% centrality bin as functions of invariant mass in Pb+Pb collisions at $\sqrt{s_{NN}} = 5.02~\mathrm{TeV}$. Results are shown for different suppression factors. The Drell--Yan dilepton contribution is also presented in the left panel as a reference.}
\label{fig:dN_v2_M}       
\end{figure}

The left panel of Fig.~\ref{fig:dN_v2_M} shows dilepton production as a function of the invariant mass in Pb+Pb collisions at $\sqrt{s_{NN}}=5.02\,\rm{TeV}$ within the 0-5\% centrality bin. The dilepton yield includes contributions from different collision stages, including pre-equilibrium, hydrodynamic (labeled thermal), and the Drell–Yan process. Here, the band in the DY dilepton calculations arises from scale uncertainties in factorization and renormalization, evaluated by varying the scales from  $0.5p_T$ to  $2.0p_T$. At lower invariant masses (LMR), dilepton production is primarily dominated by thermal dileptons from the hydrodynamic phase. This dominance arises because lower invariant mass dileptons are emitted during all stages of the collision. In the intermediate invariant mass region (IMR), the contribution from pre-equilibrium dileptons surpasses thermal dileptons, becoming the primary source of dilepton production. At the same time, dilepton production from the Drell–Yan process remains consistently lower than that from the pre-equilibrium stage throughout this intermediate mass region. This suggests that dilepton signals from the pre-equilibrium stage have a potential to be observed in this range.
Also shown is the influence of chemical composition on the dilepton production from the pre-equilibrium stage. Notably, suppressing quark population leads to the suppression of the dilepton yield, particularly at higher invariant masses. This is because higher invariant mass dileptons mostly come from the earlier stages of collision, where fewer $q$ and $\bar q$ are present in the system. Higher-order suppression factors further reduce pre-equilibrium dilepton yields, suggesting that the collision system at the pre-equilibrium stage contains relatively few fermions and requires a longer time to build up chemical equilibrium.

The right panel of Fig.~\ref{fig:dN_v2_M} shows dilepton elliptic flow $v_2^{\rm ee}\{\rm SP\}$ as a function of invariant mass. The elliptic flow calculation excludes contributions from the Drell–Yan process and considers only thermal and pre-equilibrium dileptons. The flow exhibits a single-peak structure around $M \approx 1\,{\rm GeV}$, decreasing with invariant mass moving away from this peak region. For $M < 1\,{\rm GeV}$, thermal dileptons mostly originate from the later stages of the collision and exhibit a sizeable elliptic flow.
In the region of $M>1 {\rm GeV}$, elliptic flow diminishes with increasing invariant mass due to smaller momentum anisotropy at the earlier collision stages. Additionally, considering the pre-equilibrium stage leads to further suppression in the total dilepton flow, owing to the same reasons of dilepton production causing suppression at higher invariant masses. 
However, stronger suppression factors enhance the final dilepton flow, as dileptons produced at later times make a relatively greater contribution since they carry more developed flow.

\begin{figure}[h]\label{fig:dNdpt}
\centering
\includegraphics[width=9cm,clip]{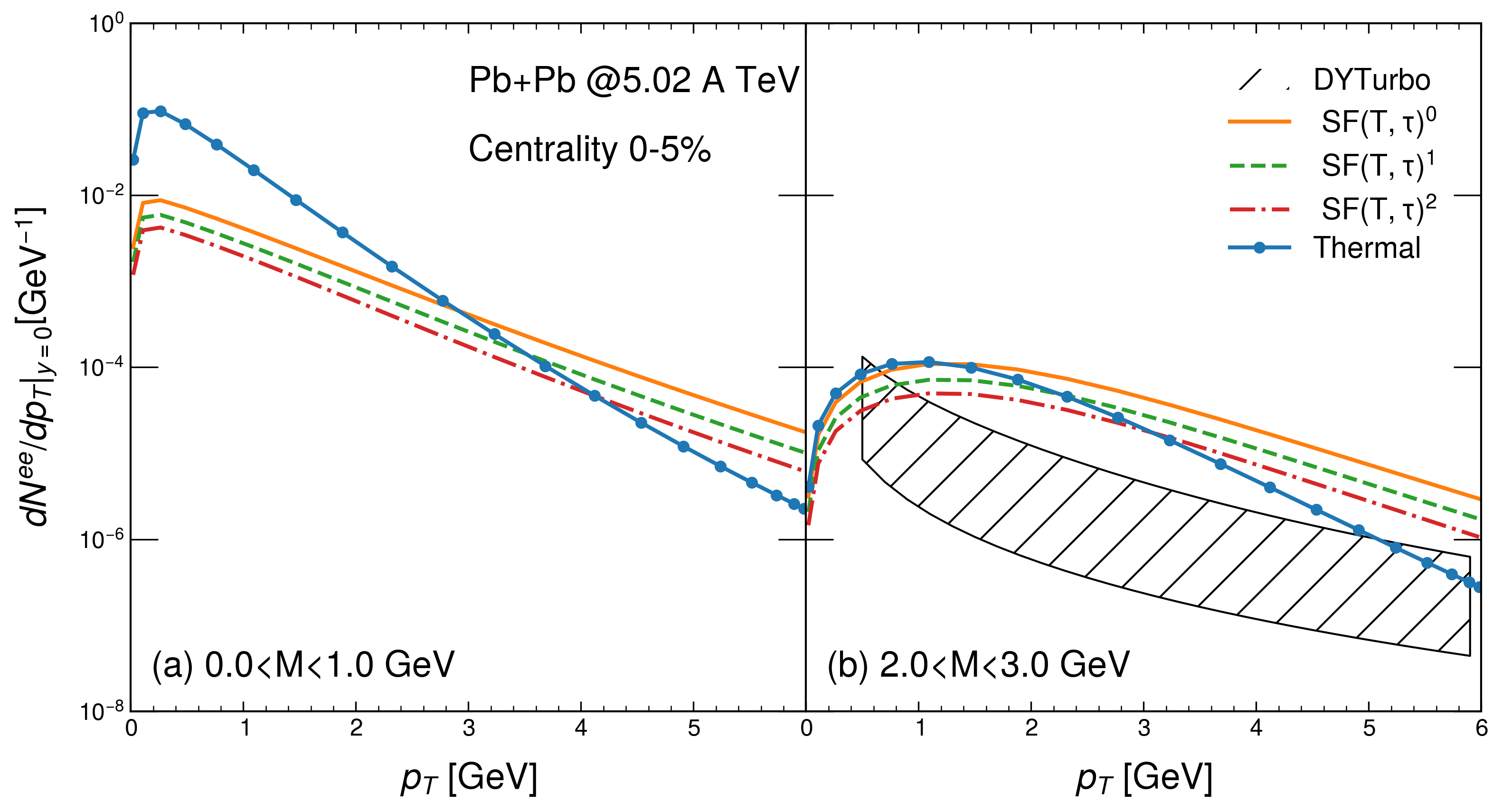}
\caption{Thermal and pre-equilibrium dilepton production with different suppression factors as a function of transverse momentum ($p_T$) in the (a) LMR: $0.0 < M < 1.0 \, \mathrm{GeV}$ and (b) IMR: $2.0 < M < 3.0 \, \mathrm{GeV}$ for 0--5\% centrality Pb+Pb collisions at $\sqrt{s_{NN}} = 5.02 \, \mathrm{TeV}$. DY dilepton contributions are included in the IMR panel.}
\label{fig:dNdpt}       
\end{figure}

In Fig.~\ref{fig:dNdpt}, we present the $p_T$-dependent dilepton production in the low invariant mass range and  intermediate invariant mass range. In the LMR, thermal dileptons dominate the low-$p_T$ region, as dileptons produced here originate primarily from the lower-temperature region of the medium evolution. Interestingly, pre-equilibrium dileptons become the dominant source at intermediate $p_T$. This indicates that dilepton production in the LMR and intermediate $p_T$ region offers an opportunity to probe the properties of the pre-equilibrium stage as well. However, it must be noted that this is an ideal scenario. In the LMR, hadronic dilepton contributions are significant and must be carefully disentangled. Until this is achieved, any interpretation of the LMR in terms of non-hadronic matter remains inconclusive.

In the IMR (panel b), because of the invariant mass range corresponding to higher-temperature regions and thus earlier stages, dilepton production exhibits a flatter slope, especially for pre-equilibrium dileptons. This behavior implies that $p_T$-dependent dilepton spectra are influenced by radial flow. Therefore, utilizing $p_T$-dependent dilepton spectra to probe radial flow at different stages of evolution is a promising area for future study.
Furthermore, the pre-equilibrium dilepton yield is comparable to the thermal dilepton yield at low $p_T$, becoming the dominant source and consistently exceeding the Drell-Yan contribution when $p_T > 1\, {\rm GeV}$. This result aligns with observations from $M$-dependent dilepton production. Hence, dileptons from the IMR provide a valuable tool for investigating pre-equilibrium dynamics, with $p_T$-dependent spectra offering further insights, particularly into radial flow.

Figure~\ref{fig:v2_pT} presents the $p_T$-dependent elliptic flow $v_2^{\rm ee}\{\rm SP\}$ of thermal and total dileptons in both the LMR and IMR. The elliptic flow of thermal dileptons exhibits a magnitude similar to that of hadrons. However, the inclusion of pre-equilibrium dileptons significantly reduces the total elliptic flow in both invariant mass regions. Furthermore, chemical equilibrium effects play a key role in the total dilepton elliptic flow. Future careful comparisons with experimental data could quantitatively constrain the chemical equilibrium during the pre-equilibrium stage.
\begin{figure}[h]\label{fig:v2_pT}
\centering
\includegraphics[width=9cm,clip]{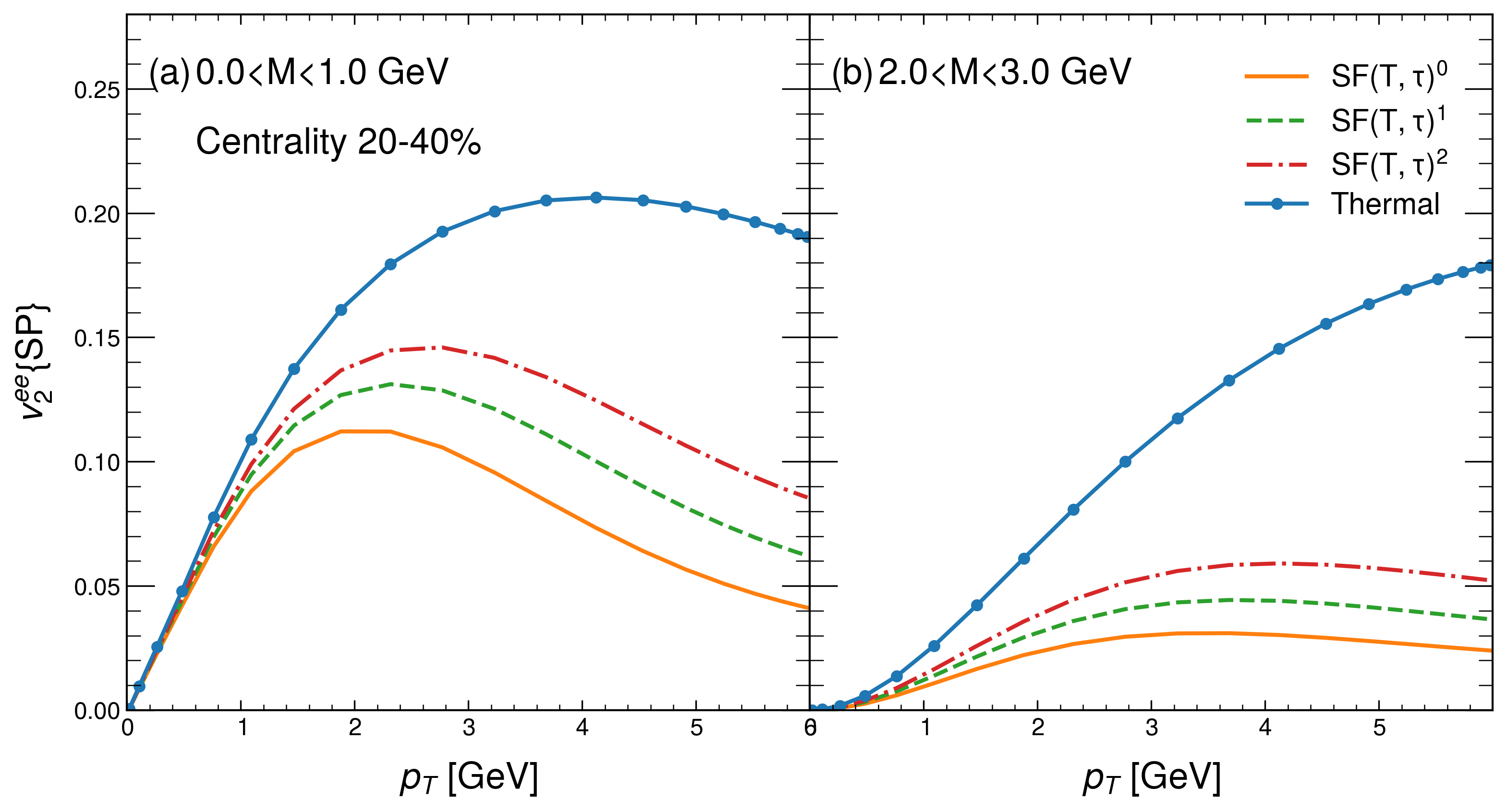}
\caption{Thermal and total dilepton elliptic flow with different suppression factors as a function of transverse momentum ($p_T$) in the (a) LMR: $0.0 < M < 1.0 \, \rm{GeV}$ and (b) IMR: $2.0 < M < 3.0 \, \mathrm{GeV}$ for 20 -- 40\% centrality Pb+Pb collisions at $\sqrt{s_{NN}} = 5.02 \, \rm{TeV}$.}
\label{fig:v2_pT}       
\end{figure}
\vspace{-1cm}
\section{Summary}
Using a framework which combines the iEBE-MUSIC hydrodynamic simulations with NLO thermal QCD dilepton emission rates, we investigate thermal dilepton production and dilepton anisotropic flow in Pb+Pb collisions at 
$\sqrt{s_{NN}}=5.02\,{\rm TeV}$ at the LHC. Our findings demonstrate that pre-equilibrium dileptons dominate the IMR, exceeding both thermal and Drell-Yan contributions. By introducing an effective suppression factor, we have performed a detailed analysis of the impact of partial chemical equilibrium on dilepton production and elliptic flow. Partial chemical equilibrium suppresses dilepton yields and enhances elliptic flow, particularly affecting flow observables. Hence, dilepton elliptic flow represents a promising observable for constraining the degree of chemical equilibrium in the pre-equilibrium stage in future studies.
\par
\vspace{0.5cm}
\noindent\textbf{Acknowledgements:} The authors thank Han Gao for many valuable discussions.  This work was supported in part by the Natural Sciences and Engineering Research Council of Canada (NSERC) [SAPIN-2024-00026 and SAPIN-2020-00048], and in part by US National Science Foundation (NSF) under grant number OAC-2004571. Computations were made on the B\'eluga super-computer system from McGill University, managed by Calcul Qu\'ebec and by the Digital Research Alliance of Canada. The operation of this supercomputer is funded by the Canada Foundation for Innovation (CFI),
Minist\`ere de l’\'Economie, des Sciences et de l’Innovation du Québec (MESI)
and le Fonds de recherche du Qu\'ebec -- Nature et technologies (FRQ-NT).
%
%

\end{document}